# Discrimination of Membrane Antigen Affinity by B cells Requires Dominance of Kinetic Proofreading over Serial Triggering

Philippos K. Tsourkas[†], Wanli Liu[§], Somkanya C Das[†], Susan K. Pierce[§], and Subhadip Raychaudhuri[*†]

[†]*Dept. of Biomedical Engineering*
*University of California, Davis*
*One Shields Avenue, Davis, CA 95616, USA*

[§]*Laboratory of Immunogenetics,*
*National Institute of Allergy and Infectious Diseases,*
*National Institutes of Health, Rockville, MD 20852*

[*]Address correspondence to: Subhadip Raychaudhuri, Dept. of Biomedical Engineering, University of California, One Shields Avenue, Davis, CA 95616, Tel: (530) 754-6716
raychaudhuri@ucdavis.edu

## ABSTRACT

B cells receptor (BCR) signaling in response to membrane-bound antigen increases with antigen affinity, a process known as affinity discrimination. We use computational modeling to show that B cell affinity discrimination requires that kinetic proofreading predominate over serial engagement. We find that if BCR molecules become signaling-capable immediately upon binding antigen, the loss in serial engagement as affinity increases results in weaker signaling with increasing affinity. A threshold time for antigen to stay bound to BCR for several seconds before the latter becomes signaling-capable, similar to kinetic proofreading, is needed to overcome the loss in serial engagement due to increasing antigen affinity, and replicate the monotonic increase in B cell signaling with affinity observed in B cell activation experiments. This finding matches well with the experimentally observed time (~ 20 seconds) required for the BCR signaling domains to undergo antigen and lipid raft-mediated conformational changes that lead to Src-family kinase recruitment. We hypothesize that the physical basis of the threshold time of antigen binding may lie in the formation timescale of BCR dimers. The latter decreases with increasing affinity, resulting in shorter threshold antigen binding times as affinity increases. Such an affinity-dependent kinetic proofreading requirement results in affinity discrimination very similar to that observed in biological experiments. B cell affinity discrimination is critical to the process of affinity maturation and the production of high affinity antibodies, and thus our results here have important implications in applications such as vaccine design.

## INTRODUCTION

The strength of B cell receptor (BCR) signaling in response to stimulation by antigen (Ag) is known to increase monotonically with the affinity of the B cell receptor for antigen, a phenomenon known as affinity discrimination.[1-8] B cell affinity discrimination has been observed starting from early membrane-proximal tyrosine phosphorylation events to late events such as lymphokine gene transcription.[3] The precise mechanisms by which B cells receptors sense antigen affinity are still the subject of current investigations.[9] While the first studies of B cell affinity discrimination focused on antigen encountered in soluble form, recent research shows that antigen presented on the surface of antigen presenting cells (APC) are potent stimulators of B cells.[4,9-20]

Further studies show that during contact between B cells and antigen presenting cells, B cells initially encounter antigen through BCR located on the B cell surface,[21] resulting in the formation of micro-clusters of 10-100 BCR/Antigen complexes.[8,12,21,22] These micro-clusters are thought to be signaling-active,[8,12,21,22] as they trigger an affinity-dependent spreading of the B cell surface over the antigen presenting cell surface, increasing the cell-cell contact area.[8] This spreading response leads to further micro-cluster formation at the leading edges,[8,21] culminating in the formation of the immunological synapse.[7,8,10,12] It has also been shown that early signaling events (~100 seconds) such as $Ca^{2+}$ flux, as well as antigen accumulation in the immunological synapse, all increase with antigen affinity.[8] Affinity discrimination has thus been observed at the earliest stages of contact between B cell receptors and antigen.

However, very little is known about how B cells discriminate between membrane antigens of varying affinity at the level of BCR-antigen micro-clusters. In this work, we use an *in silico* computational model of B cell receptor signaling to show that kinetic proofreading is the predominant mechanism by which B cell receptors discriminate membrane-antigen affinity. Originally proposed as a mechanism for how T cells discriminate between high and low affinity ligands,[23] the idea behind kinetic proofreading is that a receptor needs to undergo a series of physical modifications induced by ligand binding in order to become signaling-capable.[24-26] However, the receptor quickly reverts to its unmodified state if the ligand detaches before the fully modified state is reached. This has the net effect of setting a threshold time that the ligand needs to be bound to a receptor before the latter can become signaling-active.[24-26]

Although it bears similarities to the T cell receptor (TCR) signaling system, the BCR system also differs from it in significant ways, and what holds true for the TCR system may not be assumed to hold true for the BCR system. First, the affinity range over which B cells recognize antigen ($K_A$=$10^6$ -$10^{10}$ $M^{-1}$),[2,7,8] is much wider than that of T cells ($K_A$=$10^6$ -$10^8$ $M^{-1}$).[27] In T cells, studies indicate that TCR signaling is a non-monotonic function of antigen affinity, starting from $K_A$=$10^6$ $M^{-1}$, reaching a peak at $K_A$=$10^7$ $M^{-1}$, and decreasing thereafter (typically reaching up to $K_A$=$10^8$ $M^{-1}$).[25,26] Such non-monotonic behavior arises as a result of a competition between kinetic proofreading, which favors high affinity antigens, and serial engagement of T cell receptors by MHC peptides. Because bond life time increases with affinity, serial engagement is reduced as affinity increases.[25,26,28] Thus, if it were possible to extend such studies to affinity values above $K_A$=$10^8$ $M^{-1}$, the signaling response would continue to decrease with affinity. In contrast, the B cell signaling response increases monotonically from $K_A$=$10^6$ $M^{-1}$ to $K_A$=$10^{10}$ $M^{-1}$.[2] For very high affinity antigens, a very low dissociation rate ($k_{off}$) makes it difficult for a single antigen fragment to serially engage multiple B cell receptors. Furthermore, BCR is a bivalent molecule, whereas TCR is monovalent, and BCR is moreover expressed at much higher concentrations than TCR. This results in the BCR system having a much greater avidity than the TCR system. The question of how a B cell can discriminate between high affinity antigens is thus highly non-trivial and cannot be addressed by extrapolating what is known from TCR studies.

Our model's results show that a monotonically increasing B cell response at the level of BCR-antigen micro-clusters requires that an antigen fragment be bound to a BCR molecule for a threshold time of 5-10 seconds before that BCR's signaling domains become signaling-active, in a manner similar to kinetic proofreading. This matches well with the experimentally determined timescale of antigen and lipid-raft induced conformational changes to BCR's signaling domains and Src-family kinase recruitment.[21,22,29,30] We know that kinetic proofreading and serial engagement are generic contributing factors to receptor-ligand dynamics, but their relative contributions will depend on details of the system. Here, we show two things: First, the kinetic proofreading requirement needs to be strong enough to overcome the competing effect of serial triggering if B cells are to discriminate between antigen affinities as high as $K_A=10^8$ $M^{-1}$ and $K_A=10^9$ $M^{-1}$, and second, the timescale of the kinetic proofreading requirement that our model predicts is required for monotonically increasing B cell signaling coincides with experimentally-observed timescale for BCR association with signaling molecules such as Lyn and Syk.[21]

We find the physical basis of the threshold time requirement in the time scale of BCR dimer formation that occurs upon antigen ligation by the BCR. Our model shows that the timescale of dimer formation decreases with increasing affinity, resulting in a decreasing threshold time with increasing affinity. Such an affinity-dependent threshold time requirement results in an affinity discrimination pattern very similar to that observed in biological experiments.

## METHOD

Our technique is a Monte Carlo simulation method that builds on our previous work and has been extended to include membrane-proximal signaling events in addition to receptor-antigen binding.[31-33] Individual BCR and antigen molecules are explicitly simulated as discrete agents diffusing on virtual cell surfaces and reacting with each other subject to probabilistic parameters that directly correspond to kinetic rate constants.

### *Simulation Setup*

Because we are interested in the early stages of antigen recognition, we model a single protrusion on a B cell surface, its cytoplasmic interior, and its vertical projection onto a planar bilayer surface containing antigen. BCR is uniformly distributed over the protrusion surface, antigen over the APC surface, Lyn is anchored to the B cell protrusion surface, and Syk is distributed in the protrusion's cytoplasm. At the start of a simulation run, all of these species are distributed uniformly at random over their respective domains. At each time step, individual molecules in the population are randomly sampled to undergo either diffusion or reaction, determined by means of an unbiased coin toss.[31-33].

### *Reaction*

If a molecule has been selected to undergo reaction, we check the corresponding node on the opposing surface for a binding partner. If that is the case, a random number trial with probability $p_{\mathrm{on(i)}}$ is performed to determine if the two molecules will form a bond. BCR molecules are bivalent and can bind up to two monovalent antigen molecules, one on each Fab domain. The probability of BCR-antigen binding is denoted by $p_{\mathrm{on(BA)}}$. If a BCR/Ag complex is selected, the antigen may dissociate with probability $p_{\mathrm{off(BA)}}$ if sampled to undergo reaction. The reaction probabilities $p_{\mathrm{on(BA)}}$ and $p_{\mathrm{off(BA)}}$ are directly analogous to the kinetic rate constants $k_{\mathrm{on}}$ and $k_{\mathrm{off}}$, and their ratio, denoted as $P_{\mathrm{A}}$, is directly analogous to affinity, $K_{\mathrm{A}}$. Anchored Lyn can bind to either Ig-α or Ig-β with probability $p_{\mathrm{on(Lyn)}}$ and dissociate with probability $p_{\mathrm{off(Lyn)}}$.

We introduce a threshold antigen binding time $\mu$ such that Lyn can only bind to the Ig-α or Ig-β subunits of a BCR molecule that has bound the same antigen molecule for a length of time $\mu$. If the antigen molecule detaches before time $\mu$ is reached, the BCR molecule reverts to its basal state. Once a BCR has bound antigen for time $\mu$, however, the BCR remains signaling-capable for the duration of the simulation, even if the antigen subsequently detaches. Because at this stage we only model the very first 1-2 minutes of B cell activation, such an assumption does not conflict with subsequent internalization of BCR.[34] The length of the threshold time $\mu$ is varied in our simulations. We perform simulations where $\mu$ is a constant with respect to BCR-antigen affinity values, and simulations where the value of $\mu$ is a function of affinity $K_A$.

Lyn that is attached to either Ig-α or Ig-β may phosphorylate the Ig-α and Ig-β with probability $p_{\mathrm{phos(Ig\alpha)}}$ and $p_{\mathrm{phos(Ig\beta)}}$, respectively. Two random number trials, one with probability $p_{\mathrm{phos(Ig\alpha)}}$ and the other with probability $p_{\mathrm{phos(Ig\beta)}}$ are conducted every time an Ig-α or Ig-β subunit with Lyn attached to it is selected to undergo reaction. Syk can bind to phosphorylated Ig-α or Ig-β with probability $p_{\mathrm{on(Syk)}}$ and detach which probability $p_{\mathrm{off(Syk)}}$. A Syk molecule that is attached to phosphorylated Ig-α or Ig-β may in turn become phosphorylated with probability $p_{\mathrm{phos(Syk)}}$. The phosphorylation trial is carried out every time an Ig-α or Ig-β with an attached Syk molecule is selected for reaction. A schematic of our simplified model of membrane-proximal B cell signaling is shown in Fig. 1.

There are a total of 30 possible reactions (all reversible, and not including phosphorylation reactions) and 18 possible species (e.g. free BCR, free Ag, BCR/Ag, BCR/Ag/Lyn, BCR/$Ag_2$ BCR/$Ag_2$/Lyn, BCR/$Ag_2$/Lyn/Lyn, BCR/$Ag_2$/Lyn/Syk, etc…, not including phosphorylation status). For BCR-antigen binding, $p_{on}$ and $p_{off}$ vary with the local vertical separation between the B cell surface and the bilayer, $z$, in accordance with the linear spring model (31-33,35,36), while they are uniform for Lyn and Syk binding to Ig-α or Ig-β.

*Diffusion*

If a molecule has been selected to undergo diffusion, a random number trial with probability $p_{diff(i)}$ is used to determine whether the diffusion move will occur. The diffusion probability $p_{diff}$ is directly analogous to the diffusion coefficient $D$. The probability of diffusion of free molecules is denoted by $p_{diff(F)}$, and that of receptor-ligand complexes by $p_{diff(C)}$. If the trial with probability $p_{diff(i)}$ is successful, a direction is selected at random (4 possibilities for surface-bound species, 6 possibilities for Syk) and the appropriate neighboring nodes in that direction are checked for occupancy. Molecules may only diffuse if all the required neighboring nodes are vacant, as no two molecules are allowed to occupy the same node. The spacing between nodes is set to 10 nm. BCR molecules, BCR-antigen complexes, and BCR signalosomes are generally assumed to be much larger than antigen fragments and occupy several nodes. BCR-antigen complexes and BCR signalosomes are generally assumed to diffuse slower than free molecules,[21] hence $p_{diff(C)}$ is an order of magnitude lower than $p_{diff(F)}$.

*Sampling and time step size*

In our algorithm, the entire molecular population is randomly sampled $M$ times for diffusion or reaction during every time step. Whether a diffusion or reaction trial will occur is determined by means of an unbiased coin toss, so that the overall sampling probability for diffusion is $0.5*p_{diff}$, and that for reaction $0.5*p_{on}$ (or $0.5*p_{off}$). The number of trials $M$ is set equal to the total number of molecules (free plus complex) present in the system at the beginning of each time step, and the simulation is run for a number of time steps $T$. A distinguishing feature of our method is a mapping between the probabilistic parameters of the Monte Carlo simulation and their physical counterparts. We do this by setting $p_{diff}$ of the fastest diffusing species in our simulation equal to 1 and matching that quantity to that species' diffusion coefficient $D$. Since the nodal spacing is fixed and known, this allows us calculate the physical length of time that one of our simulation's time steps corresponds to, which here is $10^{-3}$ seconds. Once the time step size is known, it is possible to map $p_{on}$, $p_{off}$, and their ratio $P_A$ to their respective physical counterparts, $k_{on}$, $k_{off}$, and $K_A$. A detailed description of the mapping process can be found in our previous work.[31] Such a mapping makes it possible to compare our model's results to those of physical experiments to within an order of magnitude without *a priori* setting of the simulation timescale.

*Model parameters*

The parameters used in our model are listed in Table 1. Parameter values found in the literature are given on the left side of Table 1, while the appropriately mapped forms used in our model are listed on the right side of Table 1. Parameters whose values vary during experiments (such as BCR-antigen affinity and antigen concentration) are also varied in our simulations. We vary BCR-antigen affinity by order of magnitude across the physiological range for B cells ($K_A$=$10^5$-$10^{10}$ $M^{-1}$). BCR-antigen affinity is varied exactly as in B cell activation experiments, by keeping $k_{on}$ constant and varying $k_{off}$.[7,8] The same applies for parameters for which we were not able to find measured values in the literature, such as the number of Lyn and Syk molecules ($L_0$, $S_0$), and the on and off-rates of cytoplasmic reactions such as $p_{on(Lyn)}$, $p_{off(Lyn)}$, $p_{on(Syk)}$, $p_{off(Syk)}$, $p_{phos(Ig\alpha)}$, $p_{phos(Ig\beta)}$, and $p_{phos(Syk)}$. For the purposes of obtaining ballpark values for these parameters, we have adapted the values used in modeling studies of FcεRI-mediated signaling, which bears many similarities to BCR-mediated signaling.[37-38] We have been able to find values for the $K_A$ of Syk binding to Ig-α or Ig-β,[39] and hence the ratio $p_{on(Syk)}/p_{off(Syk)}$ is kept fixed in our simulations. Parametric studies conducted to gauge the effect of parameters whose measured values we were not able to find in the literature are included as Supporting Information.

## RESULTS

*Histogram plots of the number of bound antigens show affinity discrimination as $k_{off}$ decreases*

We investigate affinity discrimination by tabulating the number of bound antigen molecules, the number of signaling-active B cell receptors (i.e. with one or more phosphorylated ITAMs, denoted as pBCR), and the number of activated (phosphorylated) Syk molecules (denoted as aSyk) at the end of a simulation run of 100 physical seconds (i.e. $10^5$ time steps). BCR-antigen affinity is varied by orders of magnitude across the physiological range, from $K_A$=$10^5$ $M^{-1}$ to $K_A$=$10^{10}$ $M^{-1}$, as is done in B cell affinity discrimination experiments.[7,8] Because our simulation is stochastic in nature, the number bound antigen, pBCR, and aSyk molecules will vary from one simulation run to another. Each run of our simulation can be thought of as an *in silico* virtual experiment involving a single B cell protrusion. Thus, we perform one hundred independent trials for each affinity value and plot the results in histograms. In Fig. 2, we plot the number of bound antigen molecules as BCR-antigen affinity increases. In line with experimental results,[8] the number of bound antigen molecules increases with BCR-antigen affinity,

*Histogram plots show affinity discrimination requires a threshold time of antigen binding*

In Fig. 3, we plot histograms of the number of pBCR (Fig. 3A-C) and aSyk (Fig. 3D-F) molecules for threshold time values of $\mu$=0, 1, and 10 seconds. In the case of pBCR, we observe that with a threshold time of $\mu$=0 (Fig. 3*A*), i.e. BCR becomes signaling-capable immediately upon binding antigen, the histogram plots move in the decreasing direction as affinity increases, indicating weaker signaling with increasing affinity. This is exactly the opposite of what B cell affinity discrimination experiments show,[8] suggesting the necessity of a threshold time of an additional mechanism for affinity discrimination to be observed. With a threshold time of $\mu$=1 second (Fig. 3B), the pBCR histograms are overlapping, with the exception of the histogram for the lowest BCR-antigen affinity value, $K_A$=$10^5$ $M^{-1}$. Thus it only is possible to distinguish between this affinity value and the rest. This result shows that a threshold time of 1 second is insufficient to produce the experimentally observed affinity discrimination pattern of B cells, except between the two lowest affinity values. In addition, the histogram for the highest affinity value, $K_A$=$10^{10}$ $M^{-1}$, shows the maximum number of pBCR and aSyk does not occur at this affinity value, indicating non-monotonic dependence of signaling strength on affinity, something not seen in experiments. By contrast, when the threshold time is set to $\mu$=10 seconds (Fig. 3C), the histograms are well separated and show a monotonic increase with affinity. In this instance, it is possible to easily distinguish between all but the two highest affinity values, $K_A$=$10^9$ $M^{-1}$ and $K_A$=$10^{10}$ $M^{-1}$, while the number of pBCR is zero for every trial for BCR-antigen affinity $K_A$=$10^5$ $M^{-1}$. Increasing the threshold time to $\mu$=20 seconds shifts the histograms one order of magnitude to the left, i.e. the number of pBCR and aSyk is zero for $K_A$=$10^6$ $M^{-1}$, and the histogram for $K_A$=$10^7$ $M^{-1}$ is centered where the histogram for $K_A$=$10^6$ $M^{-1}$ was centered for $\mu$=10 seconds. This makes it no longer possible to distinguish between affinity values at the low end of the spectrum ($K_A$=$10^5$ $M^{-1}$ and $K_A$=$10^6$ $M^{-1}$) and sets the threshold of B cell activation to $K_A$=$10^7$ $M^{-1}$, in direct contradiction to the B cell activation threshold of $K_A$=$10^6$ $M^{-1}$ seen in B cell activation experiments.[7,8] Thus, only a threshold time of ~10 seconds reproduces the affinity discrimination pattern at the level of phosphorylated BCR ITAMs seen in B cell experiments. This finding correlates well (within the same order of magnitude) with recent FRET experiments that show that the Ig-α/β signaling subdomains undergo conformational changes that allow interaction with Syk approximately 20 seconds after the initiation of antigen binding.[21-29]

In the case of activated Syk molecules, when the threshold time is $\mu$=0 (Fig. 3D), the histograms overlap and it is impossible to distinguish affinity values, although perhaps a weakly decreasing trend can be discerned. For a threshold time of $\mu$=1 second (Fig. 3E), it only is possible to distinguish between $K_A=10^5$ $M^{-1}$ and higher affinity values. For a threshold time of $\mu$=10 seconds (Fig. 3F), however, the number of aSyk molecules increases with affinity and it is possible to easily distinguish between all but the two highest affinity values. As with pBCR, a threshold time of $\mu$=20 seconds contradicts the experimentally-determined B cell activation threshold affinity of $K_A=10^6$ $M^{-1}$. Our model thus predicts that only a threshold time of $\mu$=10 can reproduce the experimentally-observed affinity discrimination at the level of activated Syk molecules as well.

Of note is that when $\mu$=10 seconds, the number of pBCR and aSyk molecules is zero for the lowest affinity value, $K_A=10^5$ $M^{-1}$. This replicates the threshold of B cell activation of $K_A=10^6$ $M^{-1}$ seen in experiments.[2,4,7] Also of note is the difficulty in differentiating between the two highest affinity values, $K_A=10^9$ $M^{-1}$ and $K_A=10^{10}$ $M^{-1}$. This has also been observed in B cell activation experiments, and indicates the existence of a ceiling in B cell affinity maturation around $K_A=10^{10}$ $M^{-1}$.[2,4,7] The results for $\mu$=10 seconds are thus broadly in agreement with experimental investigations of B cell activation.

*Trial-averaged quantities also show affinity discrimination requires a threshold time of antigen binding*

In addition to histograms of the number of bound antigen, pBCR and aSyk molecules, we also plot the trial-averaged value of these quantities in Fig. 4. Trial-averaged quantities are important as they represent the signaling response integrated from either (a) multiple BCR-antigen micro-clusters within a single cell or (b) from a population of cells. As shown in Fig. 4A, the trial-averaged number of bound antigen increases monotonically with affinity, as expected, and does not vary with the threshold time $\mu$, as the threshold time only affects events downstream of antigen binding.

The number of pBCR, by contrast, is highly dependent on threshold time. In Fig. 4B, we observe that when the threshold time $\mu$=0, the trial-averaged number of pBCR decreases monotonically with increasing affinity. This is because in our simulations, as in experiments (7), affinity is increased by decreasing the dissociation probability $p_{\mathrm{off}}$ (analogous to the dissociation rate $k_{\mathrm{off}}$). Higher affinity thus means lower $p_{\mathrm{off}}$ and a longer bond lifetime. Long-lived bonds result in fewer encounters between BCR and antigen molecules, as most antigens stay bound to the same BCR molecule for a longer time. Since antigen is the limiting reagent, this means many BCR molecules never encounter antigen. Short-lived bonds, however, result in a rapid succession of binding and unbinding events between BCR and antigen, ensuring most BCR molecules encounter antigen at some point during the simulation. This effect, dubbed “serial engagement” or “serial triggering”,[28,41,42] is entirely dominant in the absence of kinetic proofreading ($\mu$=0), and is the reason for the observed decrease in the number of signaling-capable BCRs with increasing affinity.

By contrast, when the threshold time is set to $\mu$=10 seconds (Fig. 4B), the number of pBCR increases monotonically with affinity. This shows that kinetic proofreading is dominant at this threshold time value. As Lyn can only phosphorylate BCR molecules that have bound the same antigen molecule for 10 seconds or longer, the shorter bond lifetime associated with low affinity results in few BCR molecules that meet this criterion at low affinity, but many BCR molecules that do so at high affinity. This leads to an increase in the number of phosphorylation events, and hence in the number of pBCR and aSyk molecules, as affinity increases. Our model's reproduction of the activation affinity threshold of $K_A=10^6$ $M^{-1}$ and affinity ceiling of $K_A=10^{10}$ $M^{-1}$ with $\mu=10$ seconds can also be clearly seen in Fig. 4B. For the case of a threshold time of $\mu=1$ second, the number of pBCR varies non-monotonically with increasing affinity, indicating a competition between serial triggering and kinetic proofreading. Kinetic proofreading appears dominant at the lower end of the affinity range, while serial triggering appears to dominate at the higher end, with signaling strength reaching its peak at mid-range affinity values. Such a balance between kinetic proofreading and serial triggering leads to the non-monotonic signaling activation seen in T cells,[25,26,28] but is not seen in B cells.

The pattern in the number of aSyk molecules (Fig. 4C) follows that of pBCR for all threshold time values, as Syk activation occurs downstream of BCR ITAM phosphorylation. Taken together, our model's results indicate that B cell affinity discrimination requires a kinetic proofreading-type mechanism involving a threshold time of the order of 10 seconds.

*Time course of signaling activation*

In Fig. 5, we plot the time evolution of the number pBCR and aSyk for each order of magnitude in affinity between $K_A=10^5$ $M^{-1}$ to $K_A=10^{10}$ $M^{-1}$. Threshold time $\mu=0$ is shown in the top row (Fig. 5A and D), $\mu=1$ second in the middle row (Fig. 6B and E), and $\mu=10$ seconds in the bottom row (Fig. 5C and F). For threshold time $\mu=0$, the decrease in pBCR and aSyk with increasing affinity seen in Figs. 3A and 3D is readily observable for all times. For threshold time $\mu=1$ second, it only is possible to distinguish between $K_A=10^5$ $M^{-1}$ and the rest at all times. For threshold time $\mu=10$ seconds, the increase in pBCR with increasing affinity is observable at all times, and it is possible to distinguish among affinity values, as in Fig. 3C and 3F. The number of pBCR is zero at all times for $K_A=10^5$ $M^{-1}$ for threshold time $\mu=10$ seconds.

*Affinity-dependent threshold time*

Recent experimental studies indicate that BCR molecules become signaling capable through successive conformational changes mediated by BCR binding to membrane bound antigens and subsequent formation of BCR dimers.[21,25,43] We used our Monte Carlo simulation method to estimate the timescale of BCR dimer formation for each affinity value, and then used the timescale obtained in this fashion as the threshold time for BCR to become capable of binding Lyn. The timescale of dimmer formation decreases with increasing affinity in a non-linear manner, and the average time for BCR dimer formation as a function of affinity is shown in Table 2.

In Figure 6, we plot histograms of the number of pBCR (Fig. 6C) and aSyk molecules (Fig. 6F) using the times in Table 2 as the value of the threshold time $\mu$ for each affinity value. For comparison, the results for constant threshold time $\mu=10$ s (identical to Fig. 3C and F) are shown in Fig. 6B and E, (pBCR and aSyk, respectively). Results for constant threshold time $\mu=5$ s are shown in Fig. 6A and D (pBCR and aSyk, respectively). Comparing Fig. 6B with Fig. 6C, we see that for pBCR, affinity resolution is improved for affinity-dependent threshold time (Fig. 6C) compared to constant threshold time $\mu=10$ s (Fig. 6B), especially between $K_A=10^7$ $M^{-1}$ and $K_A=10^8$ $M^{-1}$ (black and green histograms, respectively). This is due to the fact that for affinity-dependent threshold time, the strength of the kinetic proofreading requirement increases as affinity decreases, thereby resulting in fewer BCRs that successfully fulfill the kinetic proofreading requirement. For aSyk, there is not much difference in affinity discrimination between affinity dependent and constant threshold time $\mu=10$ s (compare Figs 6E and 6F).

Affinity discrimination for both affinity dependent threshold time and constant threshold time $\mu$=10 s is much better than for constant threshold time $\mu$=5 s.

In Figure 7, we plot the mean values of each of the histograms in Figure 7. The plots of the mean values of the pBCR histograms are shown in Figure 7A, while the plots for the mean values of aSyk are shown in Figure 7B. The mean values for constant threshold time $\mu$=5 s are shown in blue, those for constant threshold time $\mu$=10 s in red, and those for affinity-dependent threshold time (Table 2) in black. The overall affinity discrimination pattern is similar for affinity-dependent and constant $\mu$=10 s threshold time: A rapid increase in the mean number of pBCR and aSyk as affinity increases at the lower range of affinity, followed by leveling off at high affinity values. For constant threshold time $\mu$=5 s, the mean number of pBCR and aSyk levels off much more sharply after $K_A=10^7$ $M^{-1}$, making it harder to distinguish between affinity values. For constant threshold time $\mu$=10 s and affinity dependent threshold time, the mean number of pBCR and a Syk for low affinity ($K_A=10^5$ $M^{-1}$) is zero, although not for constant threshold time $\mu$=5 s. For affinity dependent threshold time, this is due to the threshold time of 18.7 seconds. Lack of dimerization at low affinity could be the reason why non-specific antigens fail to generate a B cell response.

In Figure 8, we plot the mean number of pBCR (Fig. 8A-C) and aSyk (Fig. 8A-D) as a function of time for constant threshold time $\mu$=5 s (Fig. 8A and D), constant threshold time $\mu$=10 s (Fig. 8B and E), and affinity-dependent threshold time (Fig. 8C and F). The best affinity discrimination is observed for affinity dependent threshold time (for all time points), as that is when there is the most separation between affinity values. The increase in the mean number of pBCR and aSyk is much more rapid for the case of affinity dependent threshold time and threshold time $\mu$=5 s compared to threshold time $\mu$=10 s. For high affinity, the mean value of pBCR and aSyk is comparable between affinity dependent threshold time and threshold time $\mu$=5 s, however the separation between the mean values for affinity-dependent threshold time is much better than for threshold time $\mu$=5 s. Interestingly, our result that the two highest affinity values, $K_A=10^9$ $M^{-1}$ and $K_A=10^{10}$ $M^{-1}$, are indistinguishable, fits well with the experimental observations that there is a ceiling for affinity discrimination, above which increasing affinity does not result in enhanced antigen accumulation and signaling strength.[8]

## DISCUSSION

In this study, we have shown that B cell affinity discrimination as seen in experiments requires that kinetic proofreading predominate over serial triggering. Our model's results show that a monotonic increase in B cell signaling strength with increasing antigen affinity, up to an affinity value of $K_A$=$10^{10}$ $M^{-1}$, requires a kinetic-proofreading-type mechanism whereby antigen needs to stay bound to BCR for a threshold time of several seconds before the Ig-α and Ig-β subunits of BCR become signaling-active. Such a kinetic proofreading requirement is necessary if the B cell receptor is to overcome the decrease in serial engagement associated with high affinity antigens and discriminate between antigens with affinities as high as $K_A$=$10^8$ $M^{-1}$ and $K_A$=$10^9$ $M^{-1}$. The threshold time of the order of 5-10 seconds predicted by our model matches well (within the same order of magnitude) with the experimentally observed time required for BCR signaling domains to undergo antigen and lipid raft-mediated conformational changes that lead to association with Syk.[21,29]

Our model shows that if BCR molecules become signaling-capable immediately after binding antigen, the decrease in serial engagement as affinity (and bond lifetime) increases results in less BCR ITAM phosphorylation and hence weaker signaling, which is the opposite of what is observed in B cell activation experiments.[7,8] A kinetic proofreading requirement of 1 second results in competition between serial engagement and kinetic proofreading and a non-monotonic affinity discrimination pattern with increasing antigen affinity. A kinetic proofreading requirement of ~10 seconds is thus necessary to reproduce the experimentally observed monotonic increase in signaling strength with increasing antigen affinity, as well as the B cell activation threshold affinity of $K_A$=$10^6$ $M^{-1}$ and ceiling of $K_A$=$10^{10}$ $M^{-1}$. If the threshold time is increased significantly past ~10 seconds, our model's results disagree with the experimentally-observed B cell threshold activation affinity of $K_A$=$10^6$ $M^{-1}$.

It is known that BCRs form dimers immediately prior to the onset of signaling. Our results show that the timescale of dimer formation decreases as BCR-antigen affinity increases. When we used the timescale of dimer formation as the threshold time of antigen binding, the results were broadly similar to those with a constant threshold antigen binding time of 10 seconds. This is due to the increase in the timescale of dimer formation as affinity decreases, which makes it increasingly difficult to satisfy the kinetic proofreading and mitigates the effects of increasing serial engagement as affinity decreases. For the lowest affinity value, the average dimerization time was so long (~18 seconds) that the kinetic proofreading requirement was never satisfied, resulting in a total absence of signaling. A lack of dimer formation could thus explain why non-specific antigens fail to activate B cells.

Experimental studies of B cell activation show a significant change in FRET intensity between BCR cytoplasmic chains within a few seconds of BCR encountering antigen.[21,29] This suggests that a lipid-raft mediated conformational change (or a series of conformational changes) occurs in BCR upon encountering membrane antigen. What is intriguing is that the above-mentioned FRET experiments show a sharp increase, followed by a decrease, in intracellular FRET between BCR signaling domains for a time scale of the order of ~ 10-100 seconds (21). Based on this finding, Tolar et al.[21] propose a mechanism of B cell signaling by which B cell receptors undergo a series of antigen and lipid raft-meditated conformational changes to a signaling capable "open" conformation within a finite time following antigen binding. Dimerization of BCR molecules is an early event in such a series of conformational changes and thus could serve as the physical basis of the threshold time proposed in our model.

An affinity-dependent signaling response at the level of micro-clusters can be integrated (from many such clusters) inside B cells into a graded downstream response that will lead to affinity-dependent spreading of the B cell surface.[8] This will consequently lead to affinity-dependent collection of antigens in the B cell immunological synapse as BCR-antigen affinity increases.[7,8] Thus, one of the major functions of the B cell immunological synapse could be to collect antigens in an affinity-dependent manner for presentation to T cells. Such an affinity-dependent presentation of antigens to T cells can, in turn, modulate the affinity-dependent signaling in mature T cells.

Our model has the distinguishing feature that the probabilistic, dimensionless parameters it employs can be mapped onto their physical counterparts, allowing a meaningful physical interpretation of the results. A threshold time of 1000 dimensionless simulation time steps can thus be mapped into a physical time of 10 seconds, for example. The prediction of a ~10 second threshold time is not sensitive to variations in the values of parameters such as the number of antigen, Lyn and Syk molecules, Lyn and Syk on/off rate, or the phosphorylation rate of Ig-α, Ig-β, and Syk (see Supporting Information). The stochastic nature of our Monte Carlo simulation allows us to estimate the overlap in signaling response between antigen affinity values through probability distribution measures of signaling activation, such as histograms. Such consideration of stochastic effects in elucidating affinity discrimination in adaptive immune cells has not been explored in earlier studies. In B cells, stochastic recognition of a few very high affinity antigens can be key to the activation of pre-plasma cells. The use of a modeling technique that includes spatial effects is also important, as it incorporates effects such as competition between Lyn and Syk for BCR ITAMs.

The intrinsic ability of B cell receptors to discriminate among antigens of varying affinity, as reflected in the increase in the number of bound antigens with increasing affinity, is modified by membrane-proximal early signaling events in a way that enhances affinity discrimination at the lower end of the affinity range, but attenuates it at the high end while maintaining the monotonic increase in signaling strength with affinity. B cell affinity discrimination at the level of single-cell signaling and activation, as captured in the current study, is further modified in *in vivo* situations.[5] Recent experiments show that only high affinity B cells responded to antigen *in vivo*, even though the strength of their signaling was only two-fold higher than that of B cells whose affinity was several orders of magnitude lower (44). Formation of BCR dimers and early signaling events can determine the stop-or-go signal for B cells interacting with antigen presenting cells, and thus provide an additional mechanism of clonal competition.

B cell affinity discrimination is critical to the process of affinity maturation that results in the production of high affinity antibodies. This has important implications in applications such as vaccine design.[44] Although our model represents a simplified version of the B cell receptor signaling pathway, it captures the essential details of the early stages of B cell activation and sheds insight into this important immunological process.

## ACKNOWLEDGEMENTS

The authors thank Dr. Emanual Maverakis, Dr. Aaron Dinner, and Dr. Stephen Kaattari for proofreading the manuscript and offering valuable advice.

P.T. and S.R. are supported by the NIH grant AI074022

**Table 1.** Experimentally measured parameter values found in the literature and the mapped probabilistic counterparts used in our simulations.

| Experimental Parameter | Measured or Estimated Value | Simulation Parameter | Mapped Value |
|---|---|---|---|
| $K_A$ BCR-antigen | $10^6$-$10^{10}$ $M^{-1}$ ‡ [7,8] | $P_{A(BA)}$ | $10^2$-$10^6$ |
| $k_{on}$ BCR-antigen | $10^6$ $M^{-1}s^{-1}$ ‡ [7,8] | $p_{on(BA)}$ | 0.1 |
| $k_{off}$ BCR-antigen | 1-$10^{-4}$ $s^{-1}$ ‡ [7,8] | $p_{off(BA)}$ | $10^{-3}$-$10^{-7}$ |
| BCR molecules/cell | ~$10^5$ [35] | $B_0$ | 500 molecules |
| Antigen concentration | 10-100 molec./μm$^2$ ‡ [7] | $A_0$ | 20-200 molecules |
| $K_A$ Ig-α/β-Lyn | $10^6$ $M^{-1}$†‡ | $P_{A(Lyn)}$ | $10^2$ |
| $k_{on}$ Ig-α/β-Lyn | ~$10^5$ molec. $^{-1}$ $s^{-1}$ †‡ | $p_{on(Lyn)}$ | 1.0 |
| $k_{off}$ Ig-α/β-Lyn | ~10-0.1 $s^{-1}$ †‡ | $p_{off(Lyn)}$ | 0.01 |
| $K_A$ Ig-α/β-Syk | $10^6$ $M^{-1}$ [39] | $P_{A(Syk)}$ | $10^2$ |
| $k_{on}$ Ig-α/β-Syk | ~$10^5$ molec. $^{-1}$ $s^{-1}$ †‡ | $p_{on(Syk)}$ | 1.0 |
| $k_{off}$ Ig-α/β-Syk | ~10-0.1 $s^{-1}$ †‡ | $p_{off(Syk)}$ | 0.01 |
| Lyn molecules/cell | 2*$10^4$†‡ | $L_0$ | 100 |
| Syk molecules/cell | 4*$10^5$†‡ | $S_0$ | 400 |
| $k_{phos(Ig\alpha)}$ | ~100 $s^{-1}$†‡ | $p_{phos(Ig\alpha)}$ | 0.1 |
| $k_{phos(Ig\beta)}$ | ~100 $s^{-1}$†‡ | $p_{phos(Ig\beta)}$ | 0.1 |
| $k_{phos(Syk)}$ | ~100 $s^{-1}$†‡ | $p_{phos(Syk)}$ | 0.1 |
| $D_{free\ molecules}$ | 0.1 μm$^2$/s [40] | $p_{diff(F)}$ | 1.0 |
| $D_{complexes}$ | ~0.01 μm$^2$/s [21] | $p_{diff(C)}$ | 0.1 |

† Represents a ballpark value calculated from[37,38]
‡ Parametric study performed

**Table 2.** Threshold times predicted by our dimerization simulation.

| BCR-Ag $K_A$ ($M^{-1}$) | Threshold time $\mu$ (sec) |
|---|---|
| $10^5$ | 18.7 |
| $10^6$ | 7.6 |
| $10^7$ | 5.8 |
| $10^8$ | 4.6 |
| $10^9$ | 4.1 |
| $10^{10}$ | 3.8 |

## FIGURE LEGENDS

**Figure 1. Schematic of the simplified B cell receptor signaling pathway simulated in our Monte Carlo method.** Antigen may bind to BCR with probability $p_{\mathrm{on(BA)}}$ (Fig. 1A). If the same antigen molecule has stayed bound to the BCR for a threshold length of time $\mu$ (Fig. 2B), Lyn may bind to either the Ig-α or Ig-β subunit with probability $p_{\mathrm{on(Lyn)}}$ (Fig. 2C) and phosphorylate both with probability $p_{\mathrm{phos(Ig\text{-}\alpha)}}$ and $p_{\mathrm{phos(Ig\text{-}\beta)}}$, respectively (Fig. 2D). Once the Ig-α or Ig-β subunits are phosphorylated (Fig. 2E), Syk may bind to them with probability $p_{\mathrm{on(Syk)}}$ (Fig. 2F) and become phosphorylated with probability $p_{\mathrm{phos(Syk)}}$ (Fig. 2G). Syk may detach with probability $p_{\mathrm{on(Syk)}}$ (Fig. 2H) regardless of the outcome of the phosphorylation trial. Subsequent antigen binding may occur, but without any consequences as far as the phosphorylation of the Ig-α or Ig-β subunits is concerned (Fig. 2H).

**Figure 2. Histogram of the numbers of bound antigen molecules.** BCR-antigen affinity is varied by orders of magnitude across the physiological range in B cells, $K_A$=$10^5$ $M^{-1}$ to $K_A$=$10^{10}$ $M^{-1}$. Because of the probabilistic nature of our simulation, one hundred trials were performed for each affinity value. The parameter values used are those listed in the right hand side column of Table 1, simulation time is $10^5$ time steps (corresponding to $T$=100 physical seconds). The number of bound antigen generally increases with affinity, as expected.

**Figure 3. Histogram plots for the number of BCRs with phosphorylated ITAMs and activated Syk molecules.** BCRs with phosphorylated ITAMs (denoted as pBCR) are shown in Fig. 3A-C, while activated Syk molecules (denoted as aSyk) are shown in Fig 3D-F. Results for threshold time $\mu$=0 seconds are shown in Fig. 3A and D, for threshold time $\mu$=1 second in Fig. 3B and E, and for threshold time $\mu$=10 seconds in Fig. 3C and F. One hundred independent trials are performed for each affinity value. These results are taken after $T$=$10^5$ time steps (equal to 100 physical seconds), with the parameter values listed in the right side column of Table 1. It only is possible to clearly distinguish between affinity values with threshold time $\mu$=10 seconds.

**Figure 4. Plot of the mean number of bound antigen (Fig 4A), pBCR (Fig. 4B), and aSyk molecules (Fig. 4C) for the histograms of Fig. 2 and Fig. 3.** Results for threshold time $\mu$=0 seconds are shown as red squares, for $\mu$=1 second as blue squares, and for $\mu$=10 seconds as black squares. Where histograms are plotted in Fig. 3, the mean value of each of these histograms is shown here. The number of bound antigen shows no variation with threshold time, in contrast to the number of pBCR and aSyk. A monotonic increase in signaling strength with affinity is only observed with threshold time $\mu$=10 seconds.

**Figure 5. Plot of the mean number pBCR (Fig. 5A-C) and aSyk (Fig. 5D-F) as a function of time.** These results are for threshold time values $\mu$=0 (Fig. 5A and D), $\mu$=1 second (Fig. 5B and E), $\mu$=10 seconds (Fig. 5C and F). The data points for $T$=100 seconds correspond to the data points in Fig. 4.

**Figure 6. Comparison in affinity discrimination between constant threshold time and variable threshold time.** BCRs with phosphorylated ITAMs (pBCR) are shown in Fig. 6A-C, while activated Syk molecules (aSyk) are shown in Fig 6D-F. Results for constant threshold time $\mu$=5 seconds are shown in Fig. 6A and D, for constant threshold time $\mu$=10 seconds in Fig. 6B and E, while results for the threshold times given in Table 2 are shown in Fig. 6C and F. One

hundred independent trials are performed for each affinity value. These results are taken after $T=10^5$ time steps (equal to 100 physical seconds), with the remaining parameters values listed in the right hand side column of Table 1.

**Figure 7. Plot of the mean number of pBCR (Fig. 7A) and aSyk (Fig. 7B) for the histograms of Figure 6.** The results for constant threshold time $\mu=5$ seconds are shown in blue, constant threshold time $\mu=10$ seconds in red, and variable threshold time with $\mu$ taken from Table 2 in black. Where histograms are plotted in Fig. 6, the mean value of each of these histograms is shown here.

**Figure 8. Plot of the mean number pBCR (Fig. 8A-C) and aSyk (Fig. 8D-F) as a function of time.** These results are for constant threshold time $\mu=5$ seconds (Fig. 8A and D), $\mu=10$ seconds (Fig. 8B and E), and variable threshold time with $\mu$ taken from Table 2 (Fig. 8C and F). The data points for $T=100$ seconds correspond to the data points in Fig. 7.

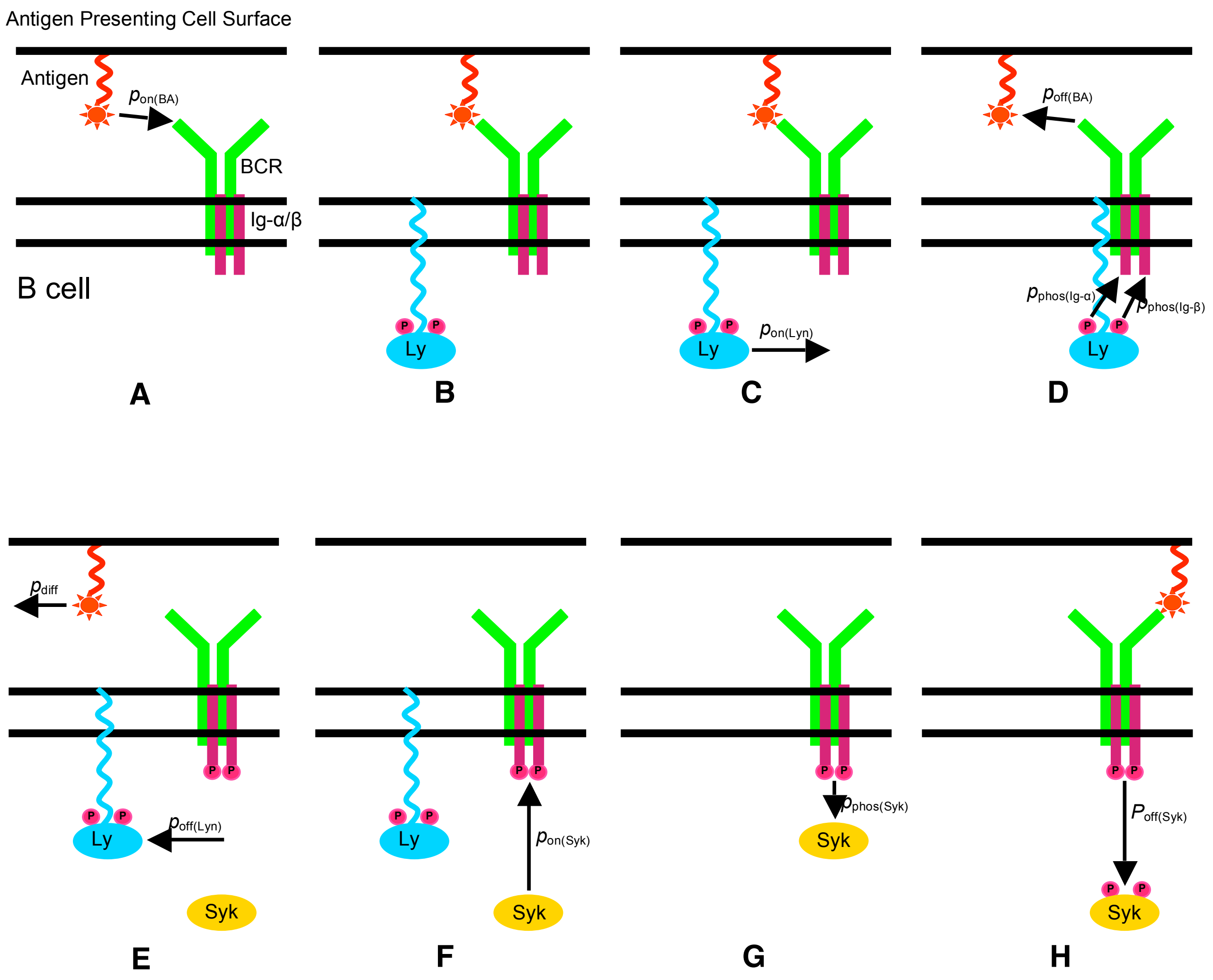

Figure 1.

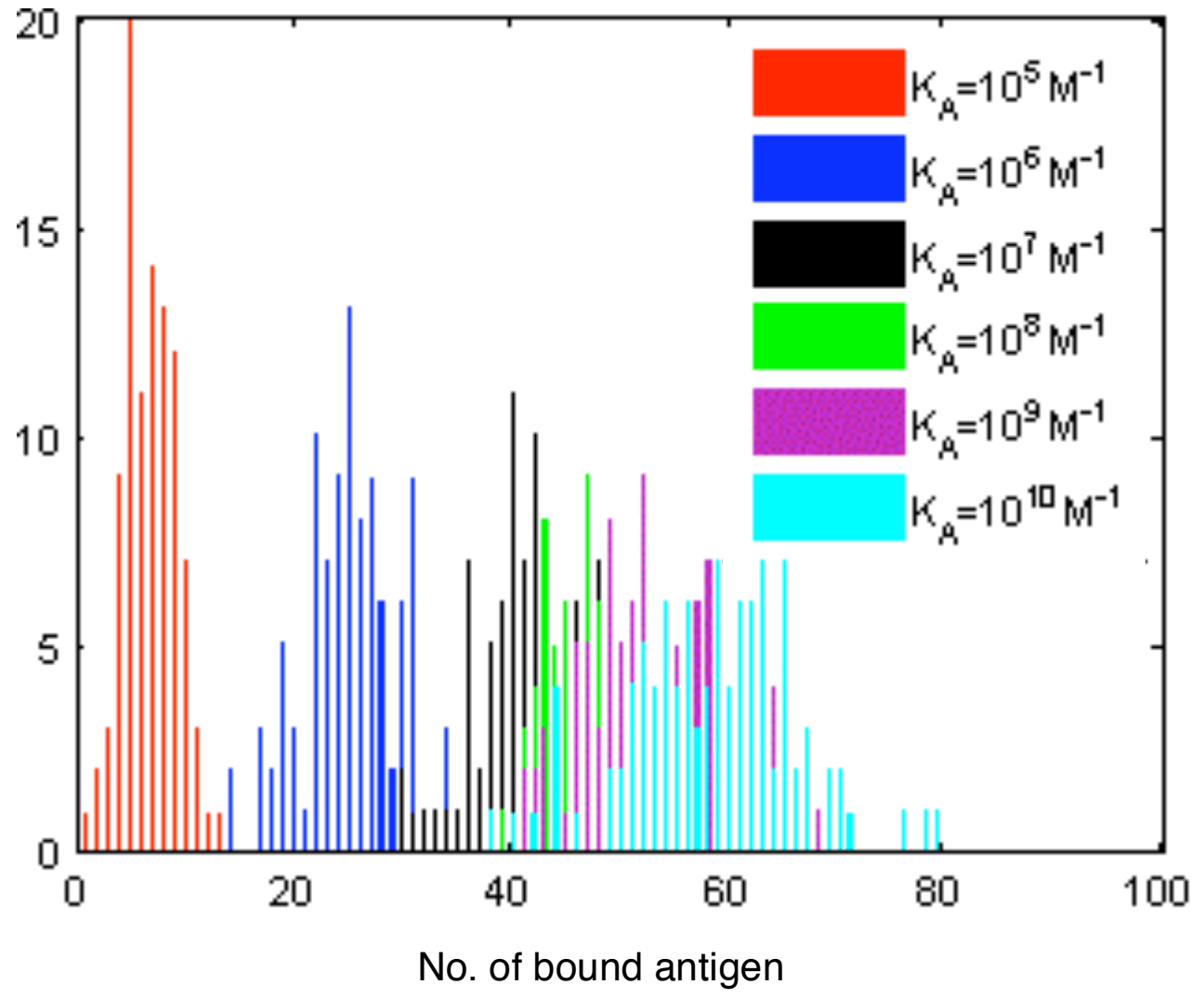

Figure 2.

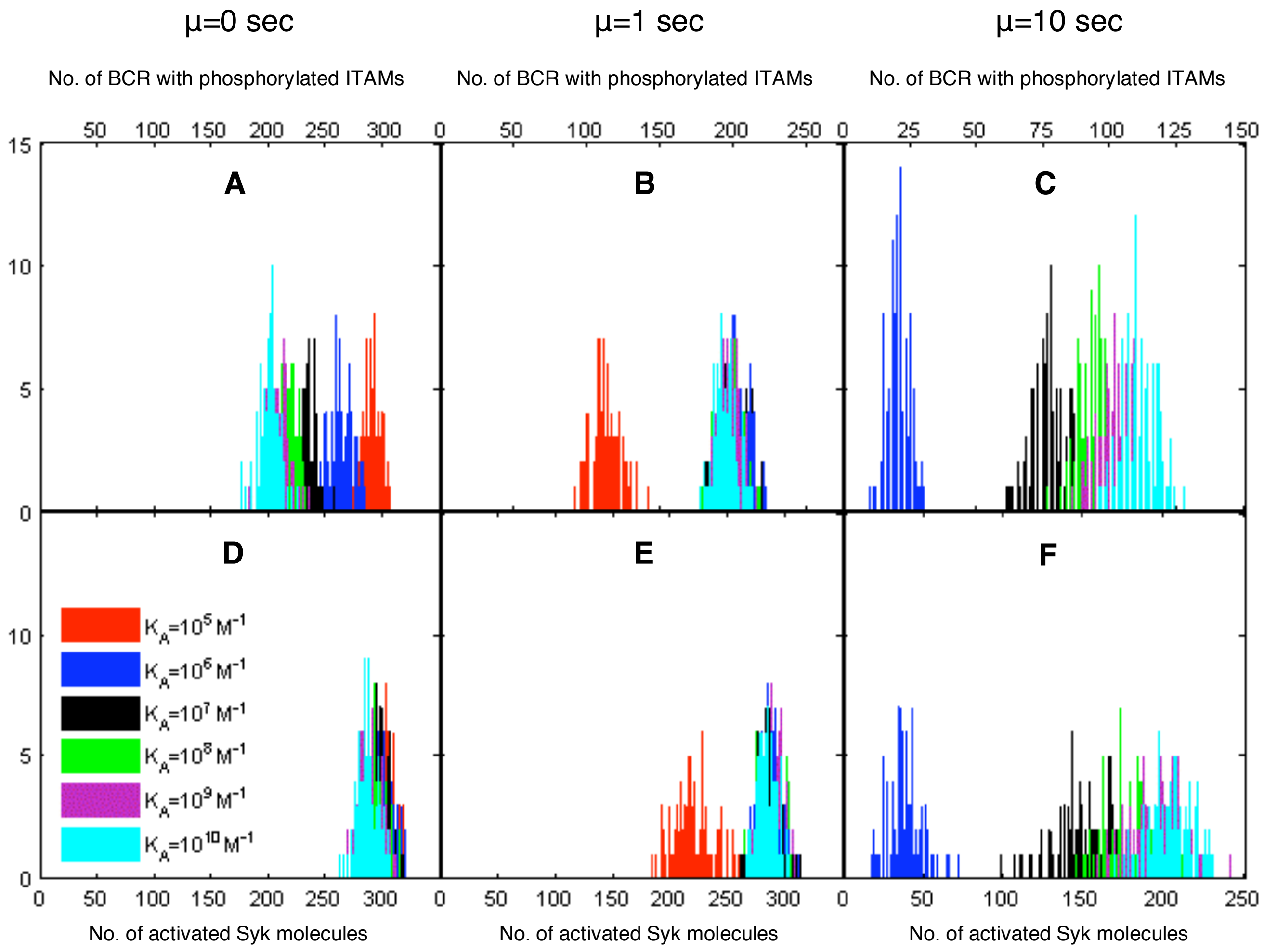
μ=0 sec
μ=1 sec
μ=10 sec
No. of BCR with phosphorylated ITAMs
No. of BCR with phosphorylated ITAMs
No. of BCR with phosphorylated ITAMs
A
B
C
D
E
F
K_A=10^5 M^-1
K_A=10^6 M^-1
K_A=10^7 M^-1
K_A=10^8 M^-1
K_A=10^9 M^-1
K_A=10^10 M^-1
No. of activated Syk molecules
No. of activated Syk molecules
No. of activated Syk molecules

Figure 3.

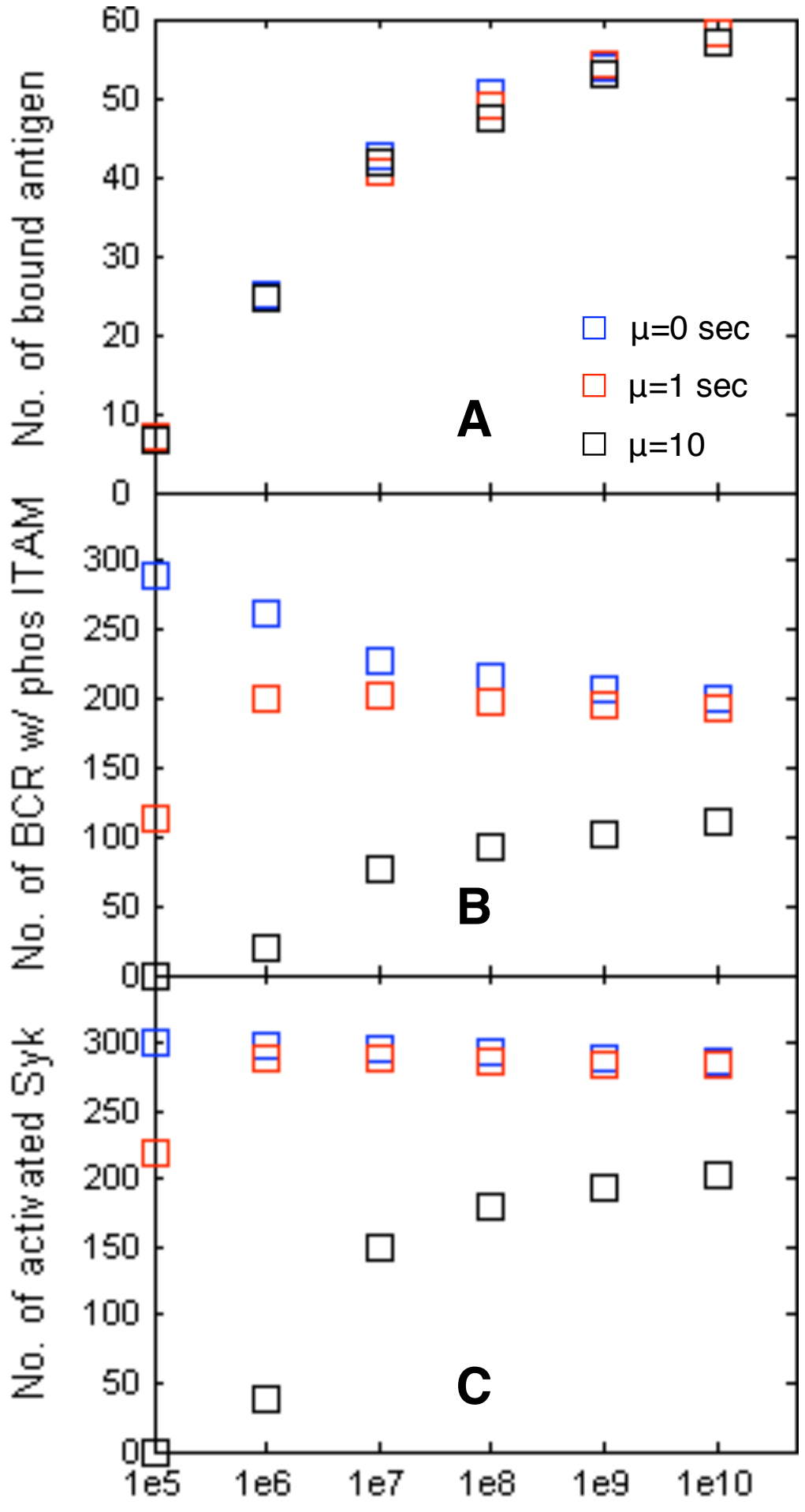

Figure 4.

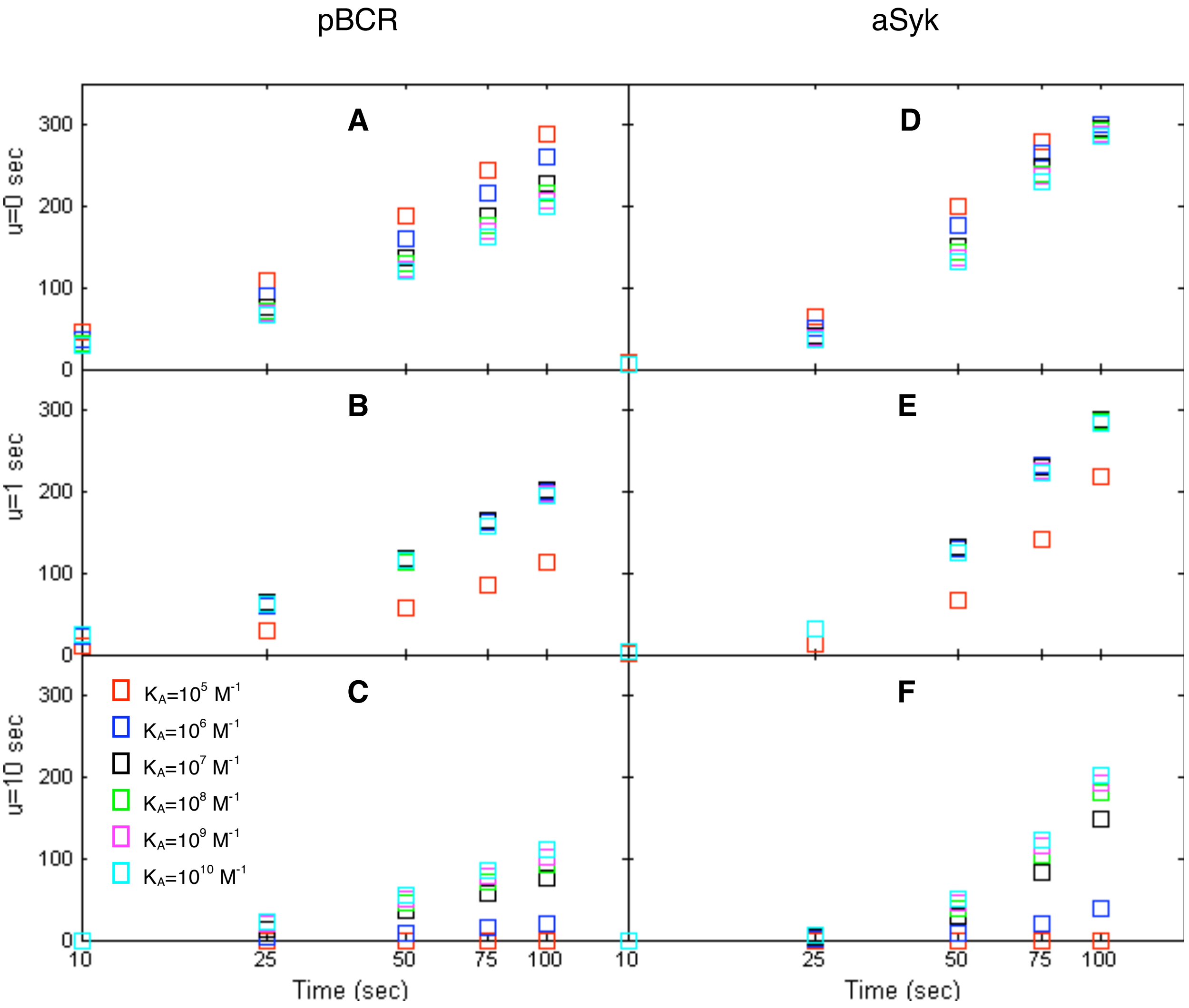

Figure 5.

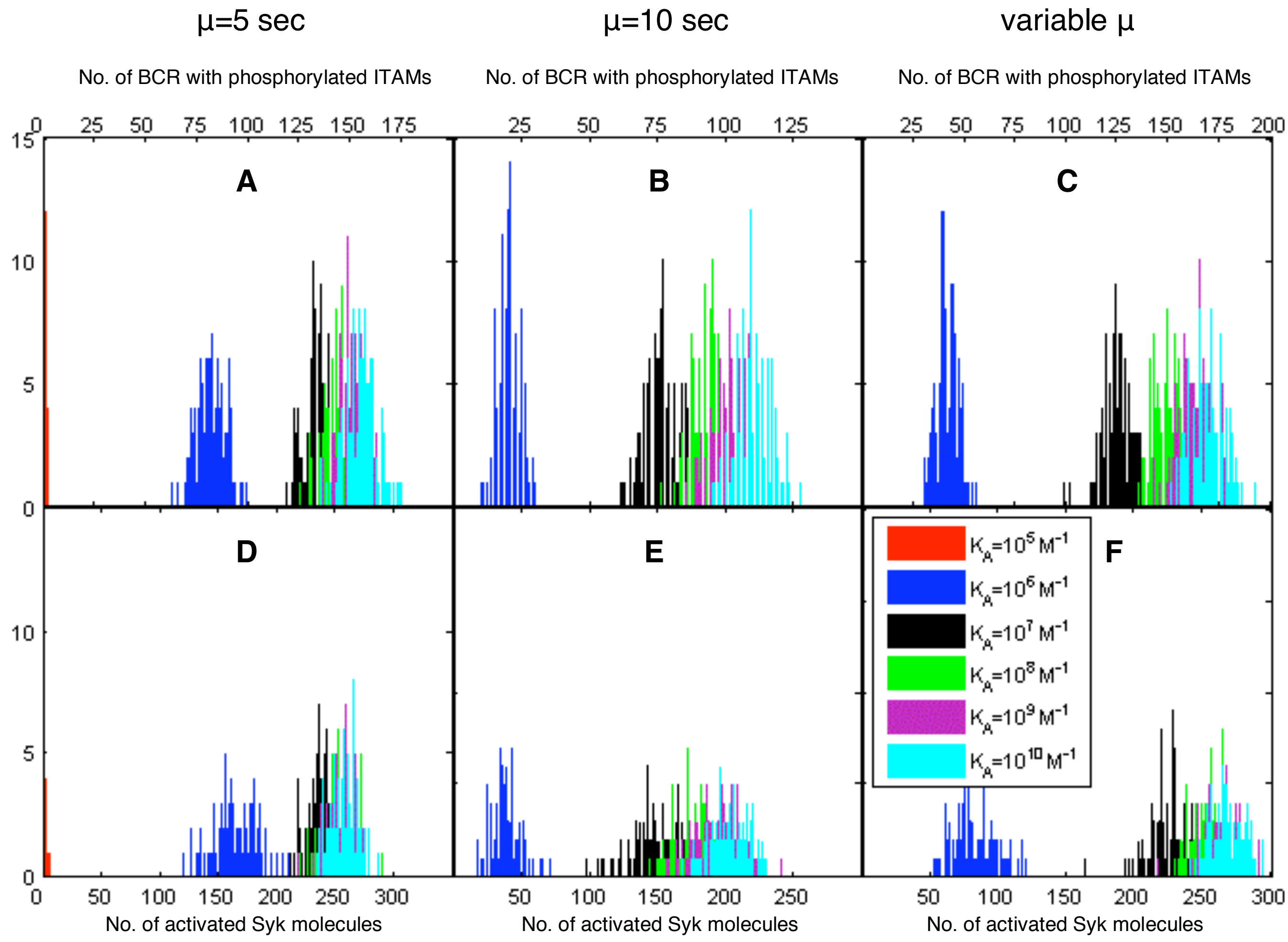

μ=5 sec
μ=10 sec
variable μ
No. of BCR with phosphorylated ITAMs
No. of BCR with phosphorylated ITAMs
No. of BCR with phosphorylated ITAMs
A
B
C
D
E
F
$K_A=10^5 M^{-1}$
$K_A=10^6 M^{-1}$
$K_A=10^7 M^{-1}$
$K_A=10^8 M^{-1}$
$K_A=10^9 M^{-1}$
$K_A=10^{10} M^{-1}$
No. of activated Syk molecules
No. of activated Syk molecules
No. of activated Syk molecules

Figure 6.

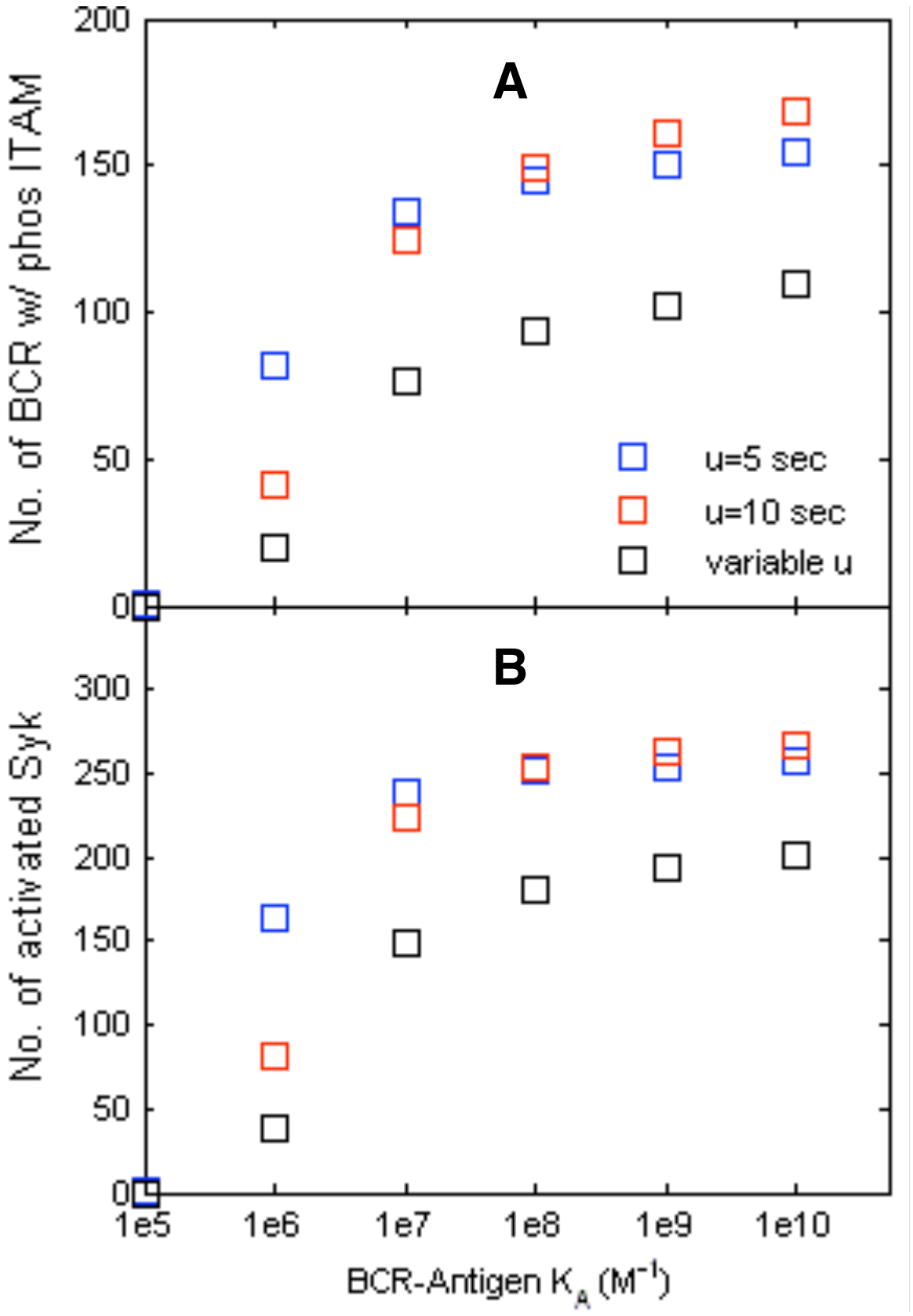

Figure 7.

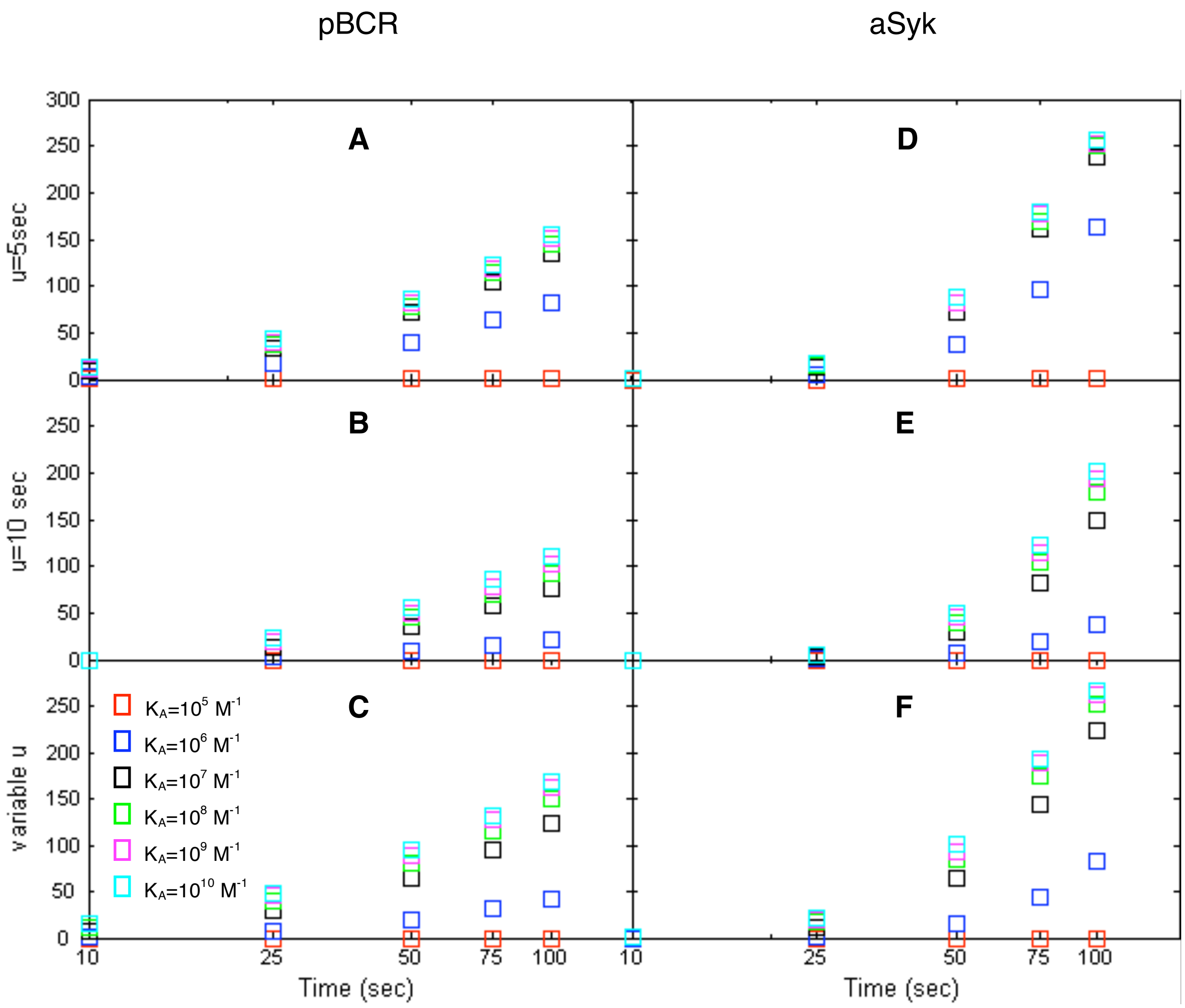

Figure 8.